\documentclass[pre,twocolumn,longbibliography,superscriptaddress]{revtex4-2}

\usepackage{amssymb}
\usepackage{amsmath}
\usepackage{amsfonts}
\usepackage{graphicx}
\usepackage{bm}
\usepackage{color}
\usepackage{multirow}
\usepackage{natbib}
\usepackage[normalem]{ulem}
\usepackage{relsize}
\usepackage{xspace}

\begin{document}
\title{Microwave spectroscopy of Schmid transition}
\author{Manuel Houzet}
\affiliation{Univ.~Grenoble Alpes, CEA, Grenoble INP, IRIG, PHELIQS, 38000 Grenoble, France}
\author{Tsuyoshi Yamamoto}
\affiliation{Faculty of Pure and Applied Physics, University of Tsukuba, Tsukuba, Ibaraki 305-8571, Japan}
\author{Leonid I. Glazman}
\affiliation{Department of Physics, Yale University, New Haven, Connecticut 06520, USA}
\date{\today}

\begin{abstract}
    {Schmid transition was introduced first as a superconductor-insulator transition in the zero-frequency response of a shunted Josephson junction in equilibrium at zero temperature. As it is typical for a quantum impurity problem, at finite frequencies the transition is broadened to a crossover. Modern attempts to find Schmid transition rely on finite-frequency measurements of a quantum circuit. We predict the frequency dependence of the admittance and reflection phase shift for a high-impedance transmission line terminated by a Josephson junction for a wide variety of devices, from a charge qubit to a transmon. Our results identify the circuit parameters allowing for the universal scaling of the responses with frequency, thus helping to identify the Schmid transition from the finite-frequency measurements.}  
\end{abstract}

\maketitle

The Schmid transition predicts that the ground-state wavefunction associated with a quantum-mechanical particle placed  in a periodic potential is either localized or extended, depending of the strength of its coupling with a dissipative environment~\cite{Schmid1983}. The existence of the transition was supported by the duality transformation found in Ref.~\cite{Schmid1983} between the two phases, and confirmed with the help of renormalization-group (RG) calculations~\cite{Bulgadaev1984,Guinea1985}. Furthermore, the RG methods allow one to argue that the transition only depends on the properties of the environment, and not on the amplitude of the periodic potential.

The particle in a periodic potential in the Schmid transition  can be associated with the phase across a Josephson junction shunted by a resistor.
If its resistance $R$ is smaller than the resistance quantum, $R<R_Q\equiv h/4e^2$, then the phase is localized in one of the minima of the Josephson potential.
Conversely, on the other side of the transition, $R>R_Q$, the phase is delocalized and the junction behaves as an insulator~\cite{Bulgadaev1984,Guinea1985}. So far, the phase diagram experimentally inferred from the dc response of shunted Josephson devices~\cite{Yagi1997,Pentilla1999} is far from reproducing the predicted phase diagram. 
 
Modern attempts to observe the Schmid transition rely on finite-frequency measurements of a superconducting quantum circuit~\cite{Murani2019,Leger2022,Kuzmin2023}, see also Ref.~\cite{Subero2022} for related heat transport measurements. 

As it is typical for a quantum impurity problem, a finite temperature or frequency broadens the quantum phase transition into a crossover. The effect of thermal fluctuations received early attention~\cite{Fisher1985,Weiss1985}. Much less is known on the role of a finite frequency that was mostly studied in perturbative regimes~\cite{Aslangul1987,Korshunov1987,Kane1992,Guinea1995,Kane1996}. A summary of some perturbative results was provided in Ref.~\cite{Kuzmin2023}.

\begin{figure}
\includegraphics[width=\columnwidth]{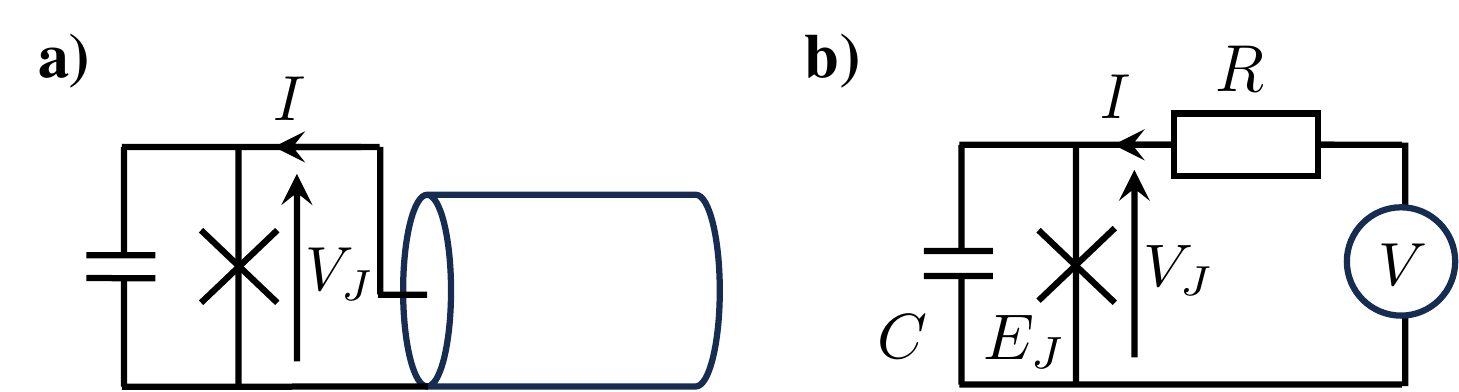}
\caption{\label{F:1} 
Two equivalent circuits: a) a transmission line terminated by a Josephson junction and b) a voltage driven resistively-shunted Josephson junction.}
\end{figure}

In this work we develop the theory of finite-frequency response functions needed for a correct interpretation of experimental data. Our results identify the circuit parameters allowing for the universal scaling of the responses with the frequency, and determine the frequency range where scaling laws apply. We predict the frequency dependence of the reflection phase shift for a high-impedance transmission line terminated by a Josephson junction, see Fig.~\ref{F:1}a, for a wide variety of devices, from a transmon ($E_J\gg E_C$) to a charge qubit ($E_J\ll E_C$).  We relate the phase shift with the admittance for the circuit depicted in Fig.~\ref{F:1}b. Here $E_J$ is the Josephson energy, and $E_C=e^2/2C$, where $C$ is the junction capacitance, is the charging energy. 

The Hamiltonian that describes a circuit formed of a Josephson junction in series with a transmission line is
\begin{equation}
\label{eq:H}
H=E_J(1-\cos\varphi)+4E_C ( N- n-{\cal N})^2+\sum_q\omega_qa^\dagger _q a_q.
\end{equation}
Here $N$ is the charge (in units of $2e$) that flows across the junction and $\varphi$ is the canonically conjugate superconducting phase difference. Furthermore, the operator that describes the charge displaced from the transmission line to the junction,
\begin{equation}
\label{eq:line-charge}
n=\frac1\pi\sum_q\sqrt{\frac{K \Delta}{\omega_q}}(a_q+a^\dagger_q),
\end{equation}
is related with the boson annihilation operator $a_q$ for a mode with energy $\omega_q=(q+\frac12)\Delta$ ($q$ positive integer) in the transmission line when it is shorted on the junction side and open on the opposite side.Here $\Delta=\pi v/L$ is the mean level spacing in a transmission line of finite length $L$, characterized by velocity $v$ and line impedance $R=R_Q/2K$ (such that the Schmid transition occurs at $K=\frac12$). The large line's capacitance, which grows linearly with its length, ensures that the zero mode not written in Eq.~\eqref{eq:line-charge} would compensate for an eventual offset charge in the electrostatic term of the Hamiltonian~\eqref{eq:H}.
To describe the circuit of Fig.~\ref{F:1}b, we take the limit $L\to \infty$ and introduce the voltage bias $V=2e \dot {\cal N}R$ with the drive variable $\cal N$.

The coupling between the junction and the line modifies the scattering properties of bosons incident from the line. In general, bosons scatter inelastically off the junction due to its nonlinearity. Still, the elastic part of the scattering matrix can be related with the circuit admittance $Y(\omega)$ at frequency $\omega$. In the one-port setup that we consider, this part reduces to the reflection amplitude $r(\omega)=e^{2i\delta(\omega)}$, with complex scattering phase $\delta(\omega)=\delta'(\omega)+i\delta''(\omega)$. Indeed, using the harmonic theory for a transmission line, we decompose the voltage and current nearby the junction in terms of incoming and outgoing waves, $V_J(\omega)=V_{\rm in}(\omega)+V_{\rm out}(\omega)$ and $I(\omega)=[V_{\rm in}(\omega)-V_{\rm out}(\omega)]/R$, respectively. The transmission line realizes an ohmic impedance, such that $I(\omega)=[V(\omega)-V_J(\omega)]/R$. Furthermore, one relates $V_{\rm out}(\omega)=r(\omega) V_{\rm in}(\omega)$. In linear response we find 
\begin{equation}
\label{eq:Y-r}
Y(\omega)\equiv \frac{I(\omega)}{V(\omega)}=\frac1{2R}\left(1-e^{2i\delta(\omega)}\right).
\end{equation}
Using the classical formula for adding impedances in series, $1/Y(\omega)=R+1/Y_{J}(\omega)$, we define the effective junction admittance,
\begin{equation}
\label{eq:r}
Y_J(\omega)=({-i}/R)\tan \delta(\omega).
\end{equation}
Note that $\delta(\omega)$ is defined modulo $\pi$; for convenience we fix it such that $0<\delta'(\omega)<\pi$. Equation~\eqref{eq:r} shows that the reflection is elastic [$\delta(\omega)$ is real] when $Y_{J}(\omega)$ is purely reactive, while the inelastic cross-section, $\sigma_{\rm in}(\omega)=1-|r(\omega)|^2$, is finite if $Y'_J(\omega)\neq 0$.

The microwave spectroscopy of a finite-length transmission line that is open on one side, such that $V_{\rm in}(\omega)=e^{-2i\omega L/v} V_{\rm out}(\omega)$, and closed by a Josephson junction on the other side, provides a direct way of measuring $\delta(\omega)$. Indeed, from the closure condition $e^{-2i\omega L/v}=e^{2i\delta(\omega)}$ we find that, when inelastic scattering is small, the frequency shift of the standing modes is $\delta\omega_n=\Delta[1/2-\delta'(\omega_n)/\pi]$, while $\sigma_{\rm in}(\omega)$ yields an internal contribution to the mode's quality factor, $Q(\omega_n)=2\pi \omega_n/\Delta \sigma_{\rm in}(\omega_n)$. This method has been implemented in a variety of experiments aiming at studying many-body physics with microwave photons in Josephson-junction arrays~\cite{Leger2019,Kuzmin2019,Kuzmin2019b,Puertas2019,Kuzmin2021,Mehta2023,Leger2022,Kuzmin2023}.

Based on these relations, one expects~\cite{Kuzmin2023} that, in the zero-frequency limit, the Schmid transition between the superconducting phase ($K>\frac12$) and the insulating phase ($K<\frac12$) manifests itself by a $\pi/ 2$ phase shift in the amplitude of wave reflection off the junction. Indeed, in the superconducting phase, the low-frequency response of the junction is inductive, such that $r=-1$ and $Y=1/R$; in the insulating phase, the low-frequency response of the junction is capacitive, such that $r=1$ and $Y=0$. 

Clearly, the zero-frequency limit is of little use for the interpretation of the microwave experiments results. On the other hand, not so much is known about the evolution with $K$ of the response at finite frequencies. Below we make specific predictions for that evolution, focusing mostly on the scaling (universal) regimes. Before doing that, we recall two simple limits, $K\gg 1$ and $K\ll 1$, respectively. Their analysis will help us to determine the domain of parameters where one may expect large variations of the phase shift with the frequency.

We first consider the classical limit, $K\gg 1$. Here 
$Y_J(\omega)=i/\omega L_J$ with Josephson inductance $L_J=1/4e^2E_J$ at any $\omega$ up to the plasma frequency, $\omega_0=\sqrt{8E_JE_C}$, except in a narrow vicinity of $\omega_0$ on the order of the plasma resonance linewidth, $2\Gamma\equiv 1/RC$. Thus $\delta(\omega)\approx \pi/2$ hardly depends on $\omega$ in a transmon. On the other, in a charge qubit $\delta(\omega)$ varies by $\sim\pi/2$, increasing with $\omega$ from $\pi/2$ to $\pi$ in the frequency range $\omega\ll\Gamma$. The increase by $\pi/2$ occurs on the scale $\omega\sim R/L_J\ll\Gamma$.

Then we consider the opposite limit of an almost disconnected Josephson junction, $K\ll 1$. Here the low-frequency response is determined by an effective capacitance $C_\star$, $Y_J(\omega)=-i\omega C_\star$, where $C_\star$ is fixed by the sensitivity of the ground state energy to an external gate voltage in a disconnected device, $K=0$~\cite{Koch2007}. In particular, in a charge qubit, such low-frequency response holds with $C_\star\approx C$ at $\omega\ll E_C$. As a result, $\delta(\omega)\approx 0$ hardly depends on the frequency if $\omega\ll \Gamma$. On the other hand, the capacitive response of a transmon holds with $C_\star=e^2/\pi^2\lambda$ if $\omega\ll \sqrt{\lambda E_J}$~\cite{SM}. Here 
\begin{equation}
\label{eq:phase-slip}
\lambda\approx\frac8{\sqrt{\pi}}\left(8{E^3_J}{E_C}\right)^{1/4}e^{-\sqrt{8E_J/E_C}}\ll \omega_0
\end{equation}
is the phase slip amplitude. As a result, $\delta(\omega)$ largely deviates from $0$ in a frequency range $\ll \omega_0$. It actually increases by $\pi/2$ as the frequency crosses over the scale $K\lambda$. Let us emphasize that this crossover is a purely single-particle, albeit nonlinear, effect and has nothing to do with many-body physics.

Overall, the above results show that the variation of the phase by $\pi/2$ occurs in opposite limits ($K\gg 1$ and $K\ll 1$) for the charge qubit and transmon, respectively. Away from these two limits, many-body effects modify this crossover and may result in a universal scaling behavior for the reflection phases. Below we will argue that the variation of $\delta(\omega)$ by $\pi/2$ in a charge qubit at $K>1/2$ is described by a complex, $K$-dependent scaling function, 
\begin{equation}
\label{eq:scaling-transmon}
\delta(\omega)=f_{\rm qb}(\omega/\Omega_\star,K),
\end{equation}
such that it incorporates inelastic scattering, as the frequency crosses over a characteristic frequency $\Omega_\star$. Correspondingly, we will determine the complex scaling function for the variation of reflection phase
\begin{equation}
\label{eq:scaling-qubit}
\delta(\omega)=f_{\rm tr}(\omega/\omega_\star,K) 
\end{equation}
from $0$ to $\pi/2$ in a transmon at $K<1/2$ with another characteristic frequency $\omega_\star$.

Let us start with the transmon coupled to a half-infinite transmission line. Starting from Eq.~\eqref{eq:H}, we find that the low-energy properties of the circuit are described by a boundary sine-Gordon Hamiltonian~\cite{SM},
\begin{eqnarray}
\label{eq:bsG}
H&=&H_0-\lambda \cos \left(2\theta(0)+2\pi {\cal N}\right), \\
H_0&=&\int_0^\infty dx\left[\frac {vK}{2\pi }(\partial_x\varphi)^2+\frac {v}{2\pi K}(\partial_x\theta)^2\right],
\nonumber
\end{eqnarray}
defined in an energy bandwidth of the order of $\omega_0$ (its precise value is beyond the accuracy of our considerations. The Hamiltonian $H_0$ is written here in terms of the canonically conjugate phase [$\varphi(x)$] and charge [$\frac1\pi\partial\theta(x)$] variables, $[\varphi(x),\frac 1\pi\partial_x \theta(x')]= \delta(x-x')$. The same Hamiltonian in the eigenmode representation is included in Eq.~\eqref{eq:H} as its last term. The charge displaced to the transmon, which determines the current operator, is $2e(n+{\cal N})$ with $n=\frac 1\pi \theta(0)$. The second term in Eq.~\eqref{eq:bsG} describes the phase slips at the Josephson junction. Using linear response and the equations of motion derived from Eq.~\eqref{eq:bsG}, we find~\cite{SM}
\begin{equation}
\label{eq:Y}
Y(\omega)=-4e^2G_{\dot n,n}=\frac 1R\left[1- {4\pi K}{\cal G}(\omega)\right]
\end{equation}
with
\begin{equation}
\label{eq:Y3}
{\cal G}(\omega)=\frac{\lambda^2}{-i\omega}\left[G_{\sin 2\pi n,\sin2\pi n}(\omega)-G_{\sin 2\pi n,\sin2\pi n}(\omega=0)\right].
\end{equation}
Here we introduced retarded Green's functions $G_{A,B}(t)=-i\theta(t)\langle[A(t),B]\rangle$ for operators $A,B$, and the last term in Eq.~\eqref{eq:Y3} arises from the relation~\cite{Kane1992,note1}
\begin{equation}
\label{eq:identity} 
\langle \cos2\pi n \rangle=-\lambda G_{\sin 2\pi n,\sin2\pi n}(\omega=0).
\end{equation}
Equations \eqref{eq:Y3} and \eqref{eq:identity} are valid at any $\lambda$.

At $K>1/2$ the second term in $H$ of Eq.~(\ref{eq:bsG}) is irrelevant. It is easy to show~\cite{footnote} that $\delta(\omega)$ remains small at any $\omega$ by using Eq.~(\ref{eq:Y-r}) and treating $\lambda$ perturbatively in Eq.~(\ref{eq:Y}).

At $K<\frac{1}{2}$, the perturbative-in-$\lambda$ result can be cast in the form
\begin{equation}
\label{eq:high-freq-asymptote}
\delta(\omega)=\frac \pi 2 +\left[\tan 2\pi K+i\right]\left(\frac{\omega_\star}{\omega}\right)^{2-4K}.
\end{equation}
The frequency-dependent correction remains small only at large frequencies, $\omega\gg \omega_\star$. Here we introduced the crossover frequency
\begin{equation}
\label{eq:crossover}
\omega_\star=\omega_0\left(\sqrt{\frac{2 K}{\Gamma(4K)}}\frac{\pi \lambda}{\omega_0}\right)^{1/(1-2K)}.
\end{equation}
below which the RG flow points towards the strong-coupling regime of the boundary sine-Gordon model~\cite{Guinea1985}.
The negative sign of $\delta'(\omega)-\frac\pi 2\propto \tan 2\pi K$ in Eq.~\eqref{eq:high-freq-asymptote} corresponds to a capacitive response with an effective $\omega$-dependent capacitance. A finite value of $\delta''(\omega)$ corresponds to a finite inelastic cross-section. Its frequency dependence reflects a quasi-elastic process~\cite{SM} similar to the one displayed by quasi-resonant photons~\cite{Houzet2020,Burshtein2021}.

In order to go beyond perturbation theory in $\lambda$ and address the low-frequency response, $\omega\ll\omega_\star$, we use a Hamiltonian dual to Eq.~\eqref{eq:bsG},
\begin{equation}
\label{eq:H-dual} 
H=H_0-\tilde \lambda \cos\varphi(0)-{\dot {\cal N}}\varphi(0).
\end{equation}
To motivate it, we note that the failure of perturbation theory at low frequency could be ascribed to the effective pinning of the charge $\theta(0)$ to multiples of $\pi$ (in the absence of the drive). The term $\propto \tilde \lambda$ in Eq.~\eqref{eq:H-dual} accounts for the {slips} induced by quantum fluctuations between those different pinned states. The precise relation of $\tilde\lambda$ to $\lambda$, which includes all the numerical factors,
\begin{equation}
\label{eq:lesage}
\frac{\pi \tilde \lambda}{\omega_0}=\frac{\Gamma(1/2K)}{2 K}\left(\frac 1{2K\Gamma(2K)}\frac{\pi \lambda}{\omega_0}\right)^{-1/2K},
\end{equation}
was found in Ref.~\cite{Fendley1995}. Overall, Eq.~\eqref{eq:H-dual} takes the same form as the Hamiltonian for a driven Josephson junction in series with a resistor. Using linear response and the equations of motion derived from Eq.~\eqref{eq:H-dual}, we may find a relation between the admittance and $\varphi(0)$-correlations functions valid at any $\tilde \lambda$. As the Josephson term in Eq.~\eqref{eq:H-dual} is irrelevant at $K<\frac12$, a perturbative-in-$\tilde \lambda$ expansion of the admittance will be valid down  to the lowest frequencies. Using Eq.~\eqref{eq:Y-r} to relate it with the frequency shift, we may express the result obtained up to $\tilde \lambda^2$ as~\cite{SM}
\begin{equation}
\label{eq:low-asymptote}
\delta(\omega)=\tilde c(1/4K)\tilde c^{1/2K}(K)\left[\tan (\pi/2K)+i\right]\left(\frac{\omega}{\omega_\star}\right)^{1/K-2}
\end{equation}
with $\tilde c(K)=8K^3\Gamma^2(2K)/\Gamma(4K)$; $\tilde c(1/2)=1$. We note that the negative sign of $\delta'(\omega)-\frac\pi 2<0$ still corresponds to a capacitive response.

Equations~(\ref{eq:high-freq-asymptote}) and (\ref{eq:low-asymptote}) extend the scaling relations, which are well-known in the context of the Kane-Fisher theory~\cite{Kane1992} for the temperature or bias dependence of the transport across an impurity in a Luttinger liquid, to the frequency dependence of the complex-valued scattering phases. The inclusion of the non-dissipative part [$\delta^\prime(\omega)$] in the response shows the need to modify one or both Eqs.~(\ref{eq:high-freq-asymptote}) and (\ref{eq:low-asymptote}) in order to consider $K<\frac{1}{3}$. 

Indeed, the amplitude of the non-dissipative term in Eq.~(\ref{eq:low-asymptote}) diverges at $K=\frac{1}{3}$, and $\delta^\prime(\omega)$ exhibits a ``super-capacitive'' response at $K<\frac{1}{3}$: the exponent $\alpha=1/K-2$ of its $\omega$-dependence {\it exceeds} the value $\alpha=1$ of a disconnected transmon. It indicates that, at $K<\frac 13$ and $\omega\ll \omega_\star$ the capacitive response originates from another irrelevant term $c_2 [\partial_x\theta(0)]^{2}$, which needs to be added to the effective low-energy Hamiltonian~\eqref{eq:H-dual}. It accounts for the quantum fluctuations of the charge $\theta(0)$ in the vicinity of a given pinned state. Such term was introduced phenomenologically in~\cite{Guinea1995,Kane1996}. The expression for the coefficient $c_2$ in terms of $K,\lambda,\omega_0$ was found as a part of series of irrelevant terms $\sum_{n=1}^\infty c_{2n}[\partial_x\theta(0)]^{2n}$ developed in~\cite{Bazhanov1997,Lesage1999}. As a result, Eq.~\eqref{eq:low-asymptote} is replaced~\cite{inelastic} with
\begin{eqnarray}
\label{eq:low-asymptote2}
\delta(\omega)&\!\!=&\!\!\frac{\omega}{\beta(K)\omega_\star}+i\tilde c(1/4K)\tilde c^{1/2K}(K)\left(\frac{\omega}{\omega_\star}\right)^{1/K-2},\\
\frac1{\beta(K)}&\!\!\!=&\!\!\!\frac1{2\sqrt{\pi}}
{\Gamma\left(\frac{1/2}{1-2K}\right)\Gamma\left(\frac{1-3K}{1-2K}\right)}
\left(\frac{\tilde c(K)}{4K^2}\right)^{\frac1{2(1-2K)}}.
\nonumber
\end{eqnarray}
Note that the effective capacitance here, $\sim 1/\beta (K)$, depends non-trivially~\cite{divergence} on $K$. Remarkably, $\beta(0)=\sqrt{2}$ allowing one to recover $C_\star$ found at $\omega/\omega_\star\ll 1$ in the isolated-transmon ($K\ll 1$) limit, see Eq.~(\ref{eq:phase-slip}). 
 
Next we notice that the non-dissipative part of the high-frequency response, Eq.~(\ref{eq:high-freq-asymptote}), runs into trouble at $K<\frac14$. The amplitude of the non-dissipative term in Eq.~(\ref{eq:high-freq-asymptote}) diverges at $K=\frac{1}{4}$, and $\delta^\prime(\omega)$ exhibits a ``super-capacitive'' response at $K<\frac{1}{4}$. The leading $1/\omega$ asymptote of $\delta^\prime(\omega)$ comes, instead, from the second term in Eq.~\eqref{eq:Y3}. By estimating $G_{\sin 2\pi n,\sin2\pi n}(\omega=0)\approx {\rm Re} \,G_{\sin 2\pi n,\sin2\pi n}(\omega_\star)$, the asymptote for the complex-valued $\delta(\omega)$ takes the form 
\begin{equation}
\label{eq:high-freq-asymptote2}
\delta(\omega)=\frac \pi 2 {\color{blue}-} \frac{\alpha(K)\omega_\star}\omega +i\left(\frac{\omega_\star}{\omega}\right)^{2-4K}
\end{equation}
with $\alpha(K)\omega_\star={4\pi K\lambda\langle \cos 2\pi n\rangle}$. We find the precise form of the $\alpha(K)$ function,
\begin{equation}
\label{alphaK}
\alpha(K)=\frac2{\sqrt{\pi}}
{\Gamma\left(\frac{\frac12-2K}{1-2K}\right)\Gamma\left(\frac{1-K}{1-2K}\right)}
\left(\frac{4K^2}{\tilde c(K)}\right)^{\frac1{2(1-2K)}},
\end{equation}
by using the exact result for $\langle \cos2\pi n\rangle$ for the boundary sine-Gordon model at $K<\frac 14$~\cite{Fateev1997}. Reassuringly, $\alpha (0)=\sqrt{2}$, so that the $K$-dependent effective capacitance extracted from the second term in Eq.~(\ref{eq:high-freq-asymptote2}) reaches at $K=0$ the value of $C_\star$ for an isolated transmon. Furthermore, the $K\to 0$ asymptote $\omega_\star\sim K\lambda$ of the crossover frequency Eq.~(\ref{eq:crossover}) agrees with the value one obtains ignoring the many-body effects for an almost-isolated transmon.

Inspecting the capacitive terms in Eqs.~(\ref{eq:high-freq-asymptote}) and (\ref{eq:high-freq-asymptote2}), we find with the help of Eq.~(\ref{alphaK}), that the amplitude of $\delta^\prime(\omega)$ diverges at $K-\frac14\to \pm 0$. The special point $K=\frac14$ corresponds to the Toulouse limit~\cite{Guinea1985b,Kane1992}, which provides an exact result covering the crossover at $\omega_\star$~\cite{Kuzmin2023,SM},
\begin{equation}
\label{eq:Toulouse}
\frac{e^{2i\delta(\omega)}+1}2=\frac{2\omega_\star}{i\pi \omega}\ln\left(1-\frac{i\pi \omega}{2\omega_\star}\right), \quad\omega_\star=\frac{\pi^2\lambda^2}{2\omega_0}.
\end{equation}
Its low-frequency asymptote matches Eq.~\eqref{eq:low-asymptote2}. In the high-frequency limit, Eq.~(\ref{eq:Toulouse}) replaces the divergence  $\propto 1/\omega|K-\frac14|$ of the $\delta'(\omega)$ terms in Eq.~\eqref{eq:high-freq-asymptote} and \eqref{eq:high-freq-asymptote2} with a non-analytical factor $\propto( \ln\omega)/\omega$.

In general, the description of the full crossover between the low- and high-frequency asymptotes of the scaling function is a difficult problem. It can however be provided for the vicinity of the Schmid transition, $K=\frac 12$. Right at the transition, the Hamiltonian \eqref{eq:bsG} can be mapped onto a tunnel Hamiltonian for free fermions~\cite{Guinea1985}. In that case, the admittance is purely real and the frequency shift is purely imaginary, both being frequency independent. A small deviation from that point, $0<\frac12-K\ll 1$, corresponds to the case of weakly repulsive fermions in the leads~\cite{Matveev1993}. Extending the theory developed in that reference to evaluate the interaction-induced corrections to the admittance and using its relation with the phase shift, we find~\cite{SM}
\begin{equation}
\label{eq:MYG}
\tan \delta(\omega)=(i-2\pi \delta K)\left(\frac{\omega}{\omega_\star}\right)^{-4\delta K},\quad  \delta K=K-\frac12,
\end{equation}
at any $\omega$. As expected, Eq.~\eqref{eq:MYG} matches the previously found asymptotes, Eqs.~\eqref{eq:high-freq-asymptote} and \eqref{eq:low-asymptote}. 

\begin{figure}
\includegraphics[width=\columnwidth]{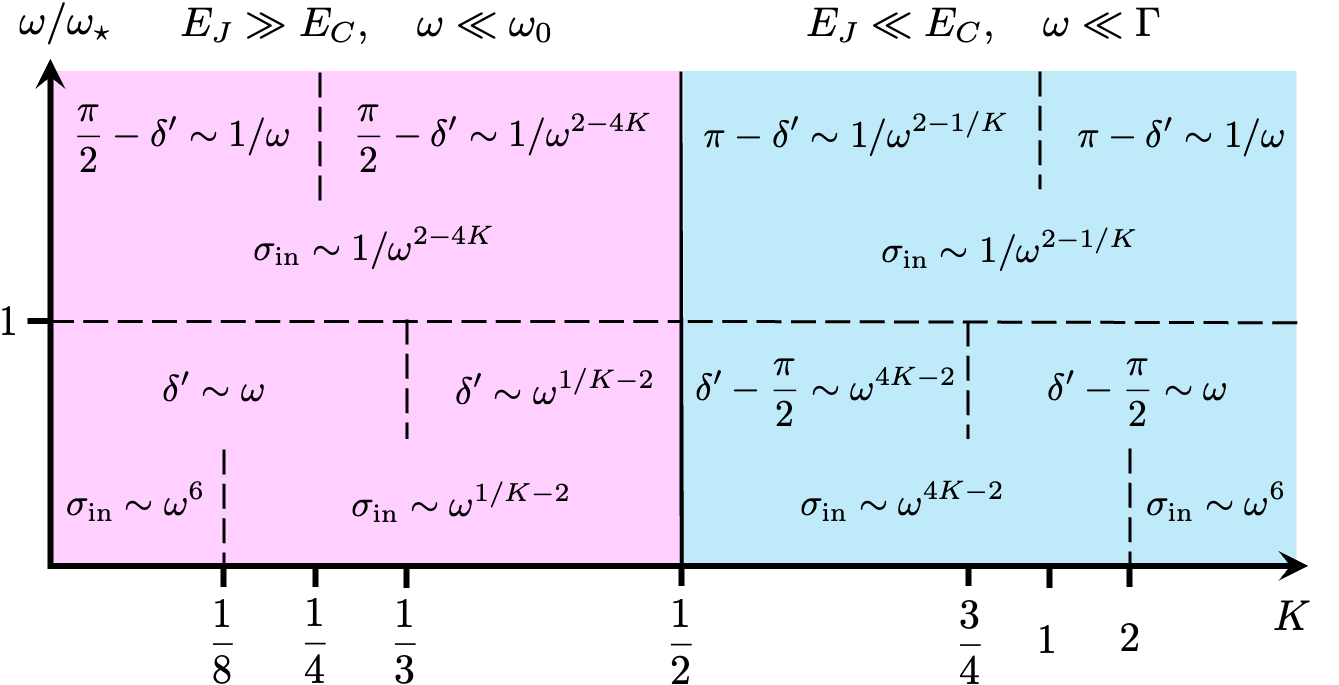}
\caption{\label{F:diag} 
High- and low-frequency asymptotes of $\delta'(\omega)$ and $\sigma_{\rm in}(\omega)$ in the scaling regime. For a transmon on the insulating side of the Schmid transition, $K<\frac12$, the results are obtained from Eqs.~\eqref{eq:high-freq-asymptote}, \eqref{eq:low-asymptote}, \eqref{eq:low-asymptote2}, \eqref{eq:high-freq-asymptote2}, and footnote~\cite{inelastic}. The asymptotes for a charge qubit on the superconducting side of the transition, $K>\frac12$, are obtained from the indicated equations using Eqs.~\eqref{eq:scaling-transmon}, \eqref{eq:scaling-qubit}, and the duality relation~\eqref{eq:duality}.}
\end{figure} 

From the asymptotes and exact results given above and illustrated in Fig.~\ref{F:diag}, we deduce that inelastic scattering, captured by $\delta''(\omega)$, provides a significant contribution to the total cross-section at $\omega$ in the vicinity of $\omega_\star$. Scattering is fully inelastic at the critical point, $K=\frac12-0$, in accordance with the exact results~\cite{Polchinski1994,Callan1994,Callan1994b} treated in the scaling limit~\cite{SM}. The appearance of a structure in $\delta(\omega)$ upon deviation of $K$ from the critical point is given by Eq.~(\ref{eq:MYG}). We remind that the observability of the scaling regime in the entire range $K<\frac12$ requires that $\omega_\star\ll\sqrt{\lambda E_J}$. As $\lambda$ varies exponentially with $E_J/E_C$, the observation of a scaling behavior in a broad dynamical range may pose a challenge for experiments.

Let us now turn to the opposite regime of the charge qubit. Starting form Hamiltonian \eqref{eq:H} at $E_J\ll E_C$, we observe that its properties at frequencies below the cutoff $\Gamma$ can be described by the same Hamiltonian \eqref{eq:H-dual} provided that one substitutes $\tilde \lambda $ with $E_J$ in it. From that duality relation, we deduce that the scaling functions in Eqs.~\eqref{eq:scaling-transmon} and \eqref{eq:scaling-qubit} are related,
\begin{equation}
\label{eq:duality}
f_{\rm qb}\left(\nu,K\right)=\frac\pi 2+f_{\rm tr}\left(\nu,\frac1{4K}\right),\quad K>\frac12,
\end{equation}
provided that one uses the proper characteristic frequency scale for the charge qubit,
\begin{equation}
\label{eq:crossover-smallEJ}
\Omega_\star=2e^\gamma\Gamma
\left(\frac1{\sqrt{2K\Gamma(1/K)}}
\frac{\pi E_J}{2e^\gamma\Gamma}\right)^{2K/(2K-1)},\quad K>\frac 12.
\end{equation}
Here the prefactor in front of the bandwidth $\Gamma$ is set by the frequency dependence of the $RC$ environment in series with the junction~\cite{WeissBook}. Furthermore, the condition $\Omega_\star\ll \Gamma$ to observe the high-frequency asymptote of the scaling function without being obscured by the classical phase shift for a capacitance, $ \tan\delta_{\rm cl}(\omega)=\omega RC$ is experimentally less stringent than for a transmon. A summary of the results is illustrated in Fig.~\ref{F:diag}.

In principle, the predictions made above could be checked numerically. Quantum Monte Carlo methods were used in \cite{Werner2005,Lukyanov2007} to evaluate the phase-phase Green function of the model in imaginary frequency. Scaling regimes were discussed there, but no attempt was made yet to perform the analytic continuation to real frequency. In Ref.~\cite{Freyn2011}, numerical RG was used to evaluate the real-frequency Green function of the boundary sine-Gordon model with an ad-hoc high-energy cutoff. It confirmed the power laws expected for the dissipative part of the admittance in the scaling regime at $\omega\ll\omega_\star$. These results cast strong doubts in the validity of a recent claim on the absence of Schmid transition, based on the numerical and functional RG~\cite{Masuki2022}. However, the latter work fails to reproduce even the simplest limit of the isolated ($K\to 0$) transmon. The validity of methods used in~\cite{Masuki2022} have been also debated~\cite{comment,reply}.

As for experiments, we note that the main argument of Refs.~\cite{Murani2019,Subero2022} for questioning the existence of the Schmid transition was the flux dependence of measured observables in devices where the Josephson junction at the end of a transmission line with $K<\frac12$ was replaced with a flux-tunable SQUID. We emphasize again that the vanishing of the effective Josephson coupling on the insulating side of the Schmid transition is only a feature of the ground state. It does not contradict the flux tunability of observables at finite frequency through the dependence of $\omega_\star$ on the bare Josephson energy, cf., e.g., Eqs.~\eqref{eq:phase-slip} and \eqref{eq:crossover} for a transmon. From this point of view, Ref.~\cite{Kuzmin2023} at least demonstrates that devices with $K<\frac12$ ($>\frac12$) show a capacitive (inductive) response, as expected from the Schmid transition paradigm. Nevertheless, the studies were performed at quite large frequency, and the task of the data analysis in the framework of scaling remains outstanding.

{\it Note added.} After finishing this work we learned about a study~\cite{Moshe} of the charge qubit limit, which was performed independently from and in parallel with our work. The results of our two studies agree with each other, wherever we were able to draw the comparison.

\begin{acknowledgments}
We thank M. Goldstein for sending us the manuscript of~\cite{Moshe} prior to making it public. MH thanks Yale Univerity for hospitality, where this work was supported by NSF Grant
No. DMR-2002275 and by ARO Grant No. W911NF-23-1-0051. TY acknowledges support from JST Moonshot R\&D–MILLENNIA Program (Grant No. JPMJMS2061). 

\end{acknowledgments}


\begin{thebibliography}{50}

\bibitem{Schmid1983}
A. Schmid, Phys. Rev. Lett. {\bf 51}, 1506 (1983).

\bibitem{Bulgadaev1984}
S. A. Bulgadaev, Pisma Zh. Eksp. Teor. Fiz. {\bf 39}, 264 (1984) [JETP Lett. {\bf 39}, 315 (1984)].

\bibitem{Guinea1985} 
F. Guinea, V. Hakim and A. Muramatsu, Phys. Rev. Lett. {\bf 54}, 263 (1985).

\bibitem{Yagi1997}
R. Yagi, S. I. Kobayashi, and Y. Ootuka,
J. Phys. Soc. Japan {\bf 66}, 3722 (1997).

\bibitem{Pentilla1999}
J. S. Penttil\"a, U. Parts, P. J. Hakonen, M. A. Paalanen, and E. B. Sonin, 
Phys. Rev. Lett. {\bf 82}, 1004 (1999).

\bibitem{Murani2019}
A. Murani, N. Bourlet, H. le Sueur, F. Portier, C. Altimiras, D. Esteve, H. Grabert, J. Stockburger, J. Ankerhold, and P. Joyez,
Phys. Rev. X {\bf 10}, 021003 (2020).


\bibitem{Leger2022}
S. L\'eger, T. S\'epulcre, D. Fraudet, O. Buisson, C. Naud, W. Hasch-Guichard, S. Florens, I. Snyman, D. M. Basko, and N. Roch,
SciPost Phys. {\bf 14}, 130 (2023).


\bibitem{Kuzmin2023}
R. Kuzmin, N. Mehta, N. Grabon, R. A. Mencia, A. Burshtein, M. Goldstein, and V. E. Manucharyan,
arXiv:2304.05806v1 [quant-ph].

\bibitem{Subero2022}
D. Subero, O. Maillet, D. S. Golubev, G. Thomas, J. T. Peltonen, B. Karimi, M. Mar\'in-Su\'arez, A. Levy Yeyati, R. S\'anchez, S. Park, and J. P. Pekola,
arXiv:2210.14953v3 [cond-mat.mes-hall].


\bibitem{Fisher1985}
M. P. A. Fisher and W. Zwerger,
Phys. Rev. B {\bf 32}, 6190 (1985).


\bibitem{Weiss1985}
U. Weiss and H. Grabert,
Phys. Lett. A {\bf 108}, 63 (1985).
%


\bibitem{Korshunov1987}
S. E. Korshunov,
Zh. Eksp. Teor. Fiz. {\bf 93},1526 (1987)
[Sov. Phys. JETP {\bf 66}, 872 (1987)]. 


\bibitem{Aslangul1987}
C. Aslangul, N. Pottier, D. Saint-James,
Journal de Physique, {\bf 48}, 1093 (1987).


\bibitem{Kane1992}
C. L. Kane and M. P. A. Fisher,
Phys. Rev. B {\bf 46}, 15233 (1992).

\bibitem{Guinea1995}
F. Guinea, G. G\'{o}mez Santos, M. Sassetti, and M. Ueda,
EPL {\bf 30}, 561 (1995).


\bibitem{Kane1996}
C. L. Kane and M. P. A. Fisher,
Phys. Rev. Lett. {\bf 76}, 3192 (1996).



\bibitem{Puertas2019}
J. Puertas Mart\'inez, S. L\'eger, N. Gheeraert, R. Dassonneville, L. Planat, F. Foroughi, Y. Krupko, O. Buisson, C. Naud, W. Hasch-Guichard, S. Florens, I. Snyman, and N. Roch,
npj Quantum Information {\bf 5}, 19 (2019).

\bibitem{Kuzmin2019}
R. Kuzmin, N. Mehta, N. Grabon, R. Mencia, and V. E. Manucharyan,
npj Quantum Information {\bf 5}, 20 (2019).

\bibitem{Leger2019}
S. L\'eger, J. Puertas-Mart\'inez, K. Bharadwaj, R. Dassonneville, J. Delaforce, F. Foroughi, V. Milchakov, L. Planat, O. Buisson, C. Naud, W. Hasch-Guichard, S. Florens, I. Snyman, and N. Roch, 
Nature Commun. {\bf 10}, 5259 (2019).

\bibitem{Kuzmin2019b}
R. Kuzmin, R. Mencia, N. Grabon, N. Mehta, Y.-H. Lin, and V. E. Manucharyan,
Nature Physics {\bf 15}, 930 (2019).

\bibitem{Kuzmin2021}
R. Kuzmin, N. Grabon, N. Mehta, A. Burshtein, M. Goldstein, M. Houzet, L. I. Glazman, and V. E. Manucharyan,
Phys. Rev. Lett. {\bf 126}, 197701 (2021).

\bibitem{Mehta2023}
N. Mehta, R. Kuzmin, C. Ciuti, and V. E. Manucharyan,
Nature {\bf 613}, {650} (2023). 


\bibitem{Koch2007}
J. Koch, T. M. Yu, J. Gambetta, A. A. Houck, D. I. Schuster,
J. Majer, A. Blais, M. H. Devoret, S. M. Girvin, and
R. J. Schoelkopf, Phys. Rev. A {\bf 76}, 042319 (2007).

\bibitem{SM}
See Supplemental Material for the details of the derivations.

\bibitem{note1}
The relation can be derived by evaluating evaluating perturbatively the response of $\langle \sin 2\pi (n+{\cal N}) \rangle$, averaged over Eq.~\eqref{eq:bsG}, to a static $\cal N$ in two different ways: either we expand the operator to be quantum-averaged in $\cal N$, or we gauge out $\cal N$ form the operator into the Hamiltonian and, then, use linear response.



\bibitem{footnote}
Matching the result with the classical limit ($K\gg 1$) requires going beyond perturbation theory in $\lambda$, see~\cite{SM}.


\bibitem{Houzet2020}
M. Houzet and L. I. Glazman,
Phys. Rev. Lett. {\bf 125}, 267701 (2020).

\bibitem{Burshtein2021}
A. Burshtein, R. Kuzmin, V. E. Manucharyan, and M. Goldstein,
Phys. Rev. Lett. {\bf 126}, 137701 (2021).


\bibitem{Fendley1995}
P. Fendley, A. W. W. Ludwig, and H. Saleur,
Phys. Rev. B {\bf 52}, 8934 (1995).

\bibitem{Bazhanov1997}
V. V. Bazhanov, S. L. Lukyanov, and A. B. Zamolodchikov,
Commun. Math. Phys. {\bf 190}, 247 (1997).



\bibitem{Lesage1999}
F. Lesage and H. Saleur,
Nuclear Physics B {\bf 546}, 585 (1999).

\bibitem{inelastic}Inelastic scattering ascribed to $\delta''(\omega)$ is given by the fractional power law in Eq.~(\ref{eq:low-asymptote2}) down to $K=\frac18$. At smaller $K$, a contribution $\propto (\omega/\omega_\star)^6$ takes over. It originates from the quartic term $\propto [\partial_x\theta(0)]^{4}$ that needs to be added to Eq.~\eqref{eq:H-dual} \cite{SM}.


\bibitem{divergence} The remaining divergence $\propto \omega /|K-\frac 13|$ of the low-frequency asymptotes \eqref{eq:low-asymptote} and \eqref{eq:low-asymptote2} likely reflects the critical behavior found in thermodynamic properties of the boundary sine-Gordon model at $K=\frac 13$~\cite{Tsvelik1995}. The frequency dependence associated with that critical behavior is currently not known.

\bibitem{Tsvelik1995}
A. M. Tsvelik,
J. Phys. A: Math. Gen. {\bf 28} L625 (1995).

\bibitem{Fateev1997}
V. Fateev, S. Lukyanov, A. Zamolodchikov, and A. Zamolodchikov,
Physics Letters B {\bf 406}, 83 (1997).

\bibitem{Guinea1985b}
F. Guinea,
Phys. Rev. B {\bf 32}, 7518 (1985).

\bibitem{Matveev1993}
K. A. Matveev, Dongxiao Yue, and L. I. Glazman,
Phys. Rev. Lett. {\bf 71}, 3351 {1993}.

\bibitem{Polchinski1994}
J. Polchinski and L. Thorlacius,
Phys. Rev. D {\bf 50}, R622(R) (1994).

\bibitem{Callan1994}
C. G. Callan and I. R. Klebanov,
Phys. Rev. Lett. {\bf 72}, 1968 (1994).

\bibitem{Callan1994b}
C. G. Callan, R. Klebanov, A. W. W. Ludwig, and J. M.Maldacena,
Nuclear Physics B {\bf 422}, 417 (1994).


\bibitem{WeissBook}
{\it Quantum dissipative systems}, U. Weiss, World Scientific (2012).


\bibitem{Werner2005}
P. Werner and M. Troyer,
Phys. Rev. Lett. {\bf 95}, 060201 (2005)

\bibitem{Lukyanov2007}
S. L. Lukyanov and P. Werner,
J. Stat. Mech. P06002 (2007).

\bibitem{Freyn2011}
A. Freyn and S. Florens,
Phys. Rev. Lett. {\bf 107}, 017201 (2011).


\bibitem{Masuki2022}
K. Masuki, H. Sudo, M. Oshikawa, and Y. Ashida,
Phys. Rev. Lett. {\bf 129}, 087001 (2022).


\bibitem{comment}
T. S\'{e}pulcre, S. Florens, and I. Snyman,
arXiv:2210.00742 [cond-mat.mes-hall].

\bibitem{reply}
K. Masuki, H. Sudo, M. Oshikawa, and Y. Ashida,
arXiv:2210.10361 [cond-mat.mes-hall].

\bibitem{Moshe} 
A. Burshtein and M. Goldstein, arXiv (2023).

\end{thebibliography}
\end{document}


\title{Supplementary Material: Microwave spectroscopy of Schmid transition}
\author{Manuel Houzet}
\affiliation{Univ.~Grenoble Alpes, CEA, Grenoble INP, IRIG, PHELIQS, 38000 Grenoble, France}
\author{Tsuyoshi Yamamoto}
\affiliation{Faculty of Pure and Applied Physics, University of Tsukuba, Tsukuba, Ibaraki 305-8571, Japan}
\author{Leonid I. Glazman}
\affiliation{Department of Physics, Yale University, New Haven, Connecticut 06520, USA}
\date{\today}

\maketitle

The supplementary material provides some details on the derivation of the results presented in the main text.

\section{Admittance of an almost isolated transmon, $K\to 0$}

In this section we show that the admittance of an isolated transmon is dominated at low frequency by the contribution of the quantum capacitance associated with it. Then, we determine the frequency shift of the modes in a transmission line weakly connected to the transmon, $K\ll 1$. Finally, we introduce a boundary sine-Gordon model that allows determining the response functions of the transmon's electromagnetic environment at any $K$. We use these results to discuss the structure of inelastic processes.

\subsection{Quantum capacitance of an isolated Josephson junction}

At $K=0$, the junction is disconnected from the transmission line. The Hamiltonian (1) in the main text reduces to 
\begin{equation}
\label{eq:Htr}
H=E_J(1-\cos\varphi)+4E_C ( N-{\cal N})^2.
\end{equation}
Here $\cal N$ is related with the current bias, $I=2e\dot {\cal N}$. The inverse admittance, $Y^{-1}_{K=0}$, is then defined as the ratio between the voltage $V_J=\langle \dot\varphi\rangle/2e$ and $I$. Using the equation of motion derived from Eq.~\eqref{eq:Htr}, $\dot \varphi=8E_C({\cal N}-N)$, and linear response, we find
\begin{equation}
\label{eq:adm-tr} 
Y^{-1}_{K=0}(\omega)=\frac 1{-i\omega C}\left[1+8 E_C G_{N,N}(\omega)\right],
\end{equation}
where $G_{N,N}(\omega)$ is the Fourier transform of the retarded Green function $G_{N,N}(t)=-i\theta(t)\langle[N(t),N]\rangle$ evaluated at ${\cal N}=0$. Using the transmon's eigenbasis of states $|s\rangle$ with energy $\varepsilon_s$, such that $\varepsilon_s<\varepsilon_{s'}$ for $0\leq s<s'$, we express Eq.~\eqref{eq:adm-tr} as
\begin{equation}
Y^{-1}_{K=0}(\omega)=\frac 1{-4e^2i\omega}\left[8E_c-\sum_{s> 0} |W_{0s}|^2\frac{2\varepsilon_{s0}}{\varepsilon_{s0}^2-\omega^2}\right]
\end{equation}
with $\varepsilon_{ss'}=\varepsilon_{s}-\varepsilon_{s'}$ and $W_{ss'}=8E_C\langle s|N|s'\rangle$. We now use an identity obtained by evaluating perturbatively $\cal N$-dependent corrections to $\varepsilon_s$~\cite{Houzet2020},
\begin{equation}
\label{eq:identity}
\frac{2e^2}{C_\star}\equiv \frac12 \frac{\partial^2\varepsilon_0}{\partial{\cal N}^2}= 4E_C-\sum_{s>0} \frac{|W_{0s}|^2}{\varepsilon_{s0}},
\end{equation}
to find
\begin{equation}
\label{eq:Ytr}
Y^{-1}_{K=0}(\omega)=\frac 1{-i\omega C_\star}-\frac{i\omega}{4e^2}\sum_{s>0} \frac{2|W_{0s}|^2}{\varepsilon_{s0}(\varepsilon_{s0}^2-\omega^2)}.
\end{equation}
The most striking feature of Eq.~\eqref{eq:Ytr} is that the low-frequency response is dominated by the contribution of the quantum capacitance $C_\star$ at any ratio $E_J/E_C$. It illustrates the insulating behavior the junction at $\omega\to 0$, in accordance with Schmid's prediction. 

In particular, in a transmon, $E_J\gg E_C$, $W_{0s}\approx -i8E_C(E_J/32E_C)^{1/4}\delta_{s,1}$ and $\varepsilon_{10}\approx \omega_0=\sqrt{8E_JE_C}$. Thus the second term in Eq.~\eqref{eq:Ytr} reduces to the classical result for an inductance $L_J=1/4e^2E_J$ in parallel with $C$. It is effectively inductive at $\omega<\omega_0$. According to Eq.~\eqref{eq:Ytr}, the quantum capacitance $C_\star$ appears in series with that effective inductance. Using 
\begin{equation}
\label{eq:tr-dispersion}
\varepsilon_0=\frac 12 \omega_0-\lambda \cos(2\pi{\cal N})
\end{equation}
 with phase-slip amplitude $\lambda$ given in Eq. (5) in the main text, we find $C_\star=e^2/\pi^2\lambda$. [Here we evaluated $C_\star$ assuming that an eventual offset charge is screened by a weakly coupled environment, see the discussion below Eq.~(2) in the main text.]

The low-frequency response is also dominated by the capacitive term in a charge qubit, $E_J\ll E_C$, where $C_\star\approx C$ and the capacitive response extends in a wide frequency range of order $ E_C$. 

\subsection{Frequency shift at $K\ll 1$}

At $K\ll 1$, we find that $Y_J\approx Y_{K=0}$  can be used to evaluate the mode frequency shift of the $q$ mode in a transmission line weakly connected to a Josephson junction. Indeed, expressing Eq.~(1) of the main text at ${\cal N}=0$ in the transmon eigenbasis, 
\begin{equation}
\label{eq:Hbis}
H=\sum_s\varepsilon_s|s\rangle\langle s|+\sum_q\omega_qa^\dagger _q a_q+4E_cn^2-n\sum_{ss'}W_{ss'}|s\rangle\langle s'|,
\end{equation}
and using perturbation theory up to order $K$, we find that $\delta\omega_q/\Delta=1/i\pi RY_{K=0}(\omega_q)$ in a wide frequency range $1/RC_\star\ll \omega_q\ll \varepsilon_{10}$ where $\delta\omega_q/\Delta\ll 1$. In particular, the correction $\delta\omega_q$ is dominated by the capacitive response, $\delta\omega_q/\Delta \approx 1/\pi R C_\star\omega_q$, in the frequency range $1/RC_\star\ll \omega_q\ll 1/\sqrt{L_J C_\star}$ in a transmon ($E_J\gg E_C$), and  $1/RC_\star\ll \omega_q\ll E_C$ in a charge qubit ($E_J\ll E_C$). These results match Eq.~(4) of the main text at large $R$ (small $K$), when $ \pi /2-\delta(\omega)\ll 1$. 

In the main text, we extrapolated the results to lower frequency, $\omega_q\ll 1/RC_\star$, using 
\begin{equation}
\label{eq:extra}
\tan \delta(\omega)=\omega R C_\star.
\end{equation} 
Equation \eqref{eq:extra} is however beyond perturbation theory in $K$. 

In the next subsection, we motivate the introduction of the boundary sine-Gordon model, Eq.~(8) in the main text, which allows extending the results at any $K$ and $\omega$.

 \subsection{Boundary sine-Gordon model}

Further insight is obtained by evaluating the ground state energy associated with the Hamiltonian~\eqref{eq:Hbis} perturbatively in $K$,
\begin{equation}
\label{eq:Eg}
E_g=\varepsilon_0+\sum_q \frac{K\Delta}{\pi^2\omega_q}\left[
4E_C-\sum_s\frac{|W_{0s}|^2}{\varepsilon_{s0}+\omega_q}\right].
\end{equation}
Here the second term yields an ${\cal N}$-dependent correction to $E_g$. Focussing on the transmon regime, we evaluate this term in logarithmic approximation after substracting Eq.~\eqref{eq:identity} to the factor between brakets. As a result, we find that the ${\cal N}$-dependent correction to $E_g$ can be accounted by the substitution 
\begin{equation}
\label{eq:subst}
\lambda\to \lambda[1-2K\ln(\omega_0/\omega_{\rm min})]
\end{equation} 
in the expression for $\varepsilon_0$, cf.~Eq.~\eqref{eq:tr-dispersion}. Here we used $\omega_{\rm min}$ to cutoff an IR divergency, while $\omega_0$ appeared as the natural cutoff for the UV divergency.

This motivates us for describing the low-frequency response of a transmon at any $K$ with a boundary sine-Gordon model [equivalent to Eq.~(8) in the main text],
\begin{equation}
\label{eq:bsG}
H_{\rm sG}=-\lambda\cos2\pi (n+{\cal N})+\sum_q\omega_qa^\dagger_qa_q,
\end{equation}
where $\omega_0$ sets the bandwidth for the boson modes (up to a numerical prefactor beyond the accuracy of our considerations). Indeed, at ${\cal N}=0$ this model yields a renormalization of the phase slip amplitude due to the environmental modes,\begin{equation}
\label{eq:renorm}
\lambda_{\rm eff}=\lambda(\omega_0/\omega_{\rm min})^{-2K},
\end{equation}
in agreement the substitution rule \eqref{eq:subst} at $K\ll 1$. The presence of $\omega_{\rm min}$ in that correction points to the fact that the term $\propto \lambda$ in Eq.~\eqref{eq:bsG} is a relevant (irrelevant) perturbation at $K<\frac 12$ ($K>\frac 12$), see the main text.

\subsection{Inelastic cross-section}

Following Ref.~\cite{Houzet2020} we also evaluate the partial inelastic cross section for a photon with frequency $\omega\ll \varepsilon_{10}$ to be converted into three photons with frequencies $\omega_1,\omega_2,\omega_3$, such that $\omega=\omega_1+\omega_2+\omega_3$
\begin{equation}
\gamma(\omega_1,\omega_2,\omega_3|\omega)=\frac{4\pi^2}{3!}\frac{K^4}{\pi^8 \omega\omega_1\omega_2\omega_3}\left(\frac{\partial^4\varepsilon_0}{\partial {\cal N}^4}\right)^2.
\end{equation}
Here we used another identity similar to Eq.~\eqref{eq:identity},
\begin{equation}
\frac1{4!}
\frac{\partial^4\varepsilon_0}{\partial {\cal N}^4}
=
\sum_{s,t>0} \frac{|W_{0s}|^2}{\varepsilon_{s0}} \frac{|W_{0t}|^2}{\varepsilon^2_{t0}}- \sum_{s,r,t>0} \frac{W_{0s}W_{sr}W_{rt}W_{t0}}{\varepsilon_{s0}\varepsilon_{r0}\varepsilon_{t0}},
\end{equation}
to simplify an equation derived from Eq.~(7) in \cite{Houzet2020} at $\omega,\omega_1,\omega_2,\omega_3\ll \varepsilon_{10}$. The contribution of the three-photon processes to the inelastic cross-section is then estimated as
\begin{eqnarray}
\gamma^{(3)}(\omega)&=&\int d\omega_1\int d\omega_2 \gamma(\omega_1,\omega_2,\omega-\omega_1-\omega_2|\omega)
\nonumber\\
&\approx& \frac{4\pi^2}{2!}\frac{2^8 K^4\lambda^2}{\omega^2}\ln^2\frac\omega{\omega_{\rm min}}.
\end{eqnarray}
Here we evaluated the integrals in logarithmic approximation. Evaluating similarly all processes where an incident photon is converted into $2m+1$ photons, we find the corresponding cross-section:
\begin{equation}
\label{eq:partial}
\gamma^{(2m+1)}(\omega)\approx \frac{4\pi^2}{(2m)!}\frac{(4K)^{2m+2}\lambda^2}{\omega^2} \ln^{2m}\frac\omega{\omega_{\rm min}}.
\end{equation}
The sum of all these processes yield the inelastic cross-section,
\begin{eqnarray}
\sigma_{\rm in}(\omega)&=&\sum_{m=1}^\infty\gamma^{(2m+1)}(\omega)
\nonumber\\
&\approx& \frac{64\pi^2K^2\lambda^2}{\omega^2}\left[\cosh\left(4K\ln\frac\omega{\omega_{\rm min}}\right)-1\right]
\nonumber\\
&\approx& \frac{32\pi^2K^2\lambda^2}{\omega^2}\left(\frac\omega{\omega_{\rm min}}\right)^{4K}
\end{eqnarray}
at $K\ln(\omega/{\omega_{\rm min}})\gg 1$. Taking into account the renormalization of $\lambda$ given by Eq.~\eqref{eq:renorm} and expression (13) in the main text for the crossover frequency at $K\ll 1$, we finally get
 \begin{equation}
 \delta''(\omega)=\frac{\sigma_{\rm in}(\omega)}4=\left(\frac{\omega_\star}\omega\right)^{2-4K},
 \end{equation}
 in agreement with Eq. (12) in the main text.

Similar to the resonance fluorescence studied in Ref.~\cite{Houzet2020}, the logarithmic factors in the partial inelastic cross-sections, Eq.~\eqref{eq:partial}, illustrate that an incident photon with frequency $\omega\gg\omega_\star$ is quasi-elastically scattered into another photon with slightly smaller frequency, as well an even number of low-frequency photons.

\section{Admittance of a transmon within the boundary sine-Gordon model}

In this section we provide details on the derivation of the circuit admittance in the transmon regime.

\subsection{Exact formula for the admittance}

\label{sec:EOM}

The Hamiltonian \eqref{eq:bsG} [Eq.~(8) of the main text] is conveniently expressed as
\begin{eqnarray}
\label{eq:BSG2}
H_{\rm sG}&=&\sum_q\frac{\omega_q}2\left(X_q^2+P_q^2\right)-\lambda\cos2\pi(n+{\cal N}),\quad
\\
\label{eq:weights} 
n&=&\frac1\pi\sum_q\sqrt{\frac{2K\Delta}{\omega_q}}X_q.
\end{eqnarray}
Here we introduced the canonically conjugate variables $X_q=(a_q+a^\dagger_q)/\sqrt{2}$ and $P_q=(a_q-a^\dagger_q)/i\sqrt{2}$.
Using linear response of the current, $I=2e(\dot n+\dot{\cal N})$, to the applied bias, $V=2eR\dot {\cal N}$, we express the admittance as
\begin{equation}
\label{eq:adm}
Y(\omega)=\frac 1R\left[1+2\pi \lambda G_{n,\sin2\pi n}(\omega)\right],
\end{equation}
where the Green function is evaluated with Hamiltonian \eqref{eq:BSG2} at ${\cal N}=0$, and we used $G_{\dot n,\sin2\pi n}(\omega)=-i\omega G_{n,\sin2\pi n}(\omega)$.
Then we derive equivalent formulas for $Y(\omega)$ by using identities between various Green functions. Indeed, using the equations of motion
\begin{equation}
\dot X_q=\omega_qP_q, \quad
\dot P_q=-\omega_qX_q-2\lambda\sqrt{\frac{2K\Delta}{\omega_q}}\sin 2\pi n,
\end{equation}
we find
\begin{equation}
G_{X_q,B}(\omega)=\frac{\sqrt{2K\Delta\omega_q}}{\omega^2-\omega_q^2}\left[\frac 1\pi\left\langle \frac{\partial B}{\partial n}\right\rangle+2\lambda G_{\sin2\pi n,B}(\omega)\right]
\end{equation}
for an arbitrary operator $B$ expressed as a function of $n$ only. Summing Green functions with weights given by Eq.~\eqref{eq:weights} we find
\begin{equation}
\label{eq:id}
G_{n,B}(\omega)=\frac K{i\omega}\left[\frac 1\pi\left\langle \frac{\partial B}{\partial n}\right\rangle+2\lambda G_{\sin2\pi n,B}(\omega)\right],
\end{equation}
where we evaluated $\sum_q{2K\Delta}/({\omega^2-\omega_q^2})=\pi K/{i \omega}$ in the limit $\Delta\to 0$. (Recall that we consider retarded Green functions, such that $\omega$ contains a small, positive imaginary part.) Using $B=n$ and $G_{n,\sin 2\pi n}=G_{\sin 2\pi n,n}$, and inserting the corresponding Eq.~\eqref{eq:id} into Eq.~\eqref{eq:adm}, we find $Y(\omega)=-4e^2 G_{\dot n, n}(\omega)$, in agreement with Eq.~(9) in the main text. Using $B=\sin 2\pi n$ and footnote [25] in the main text, we also find 
\begin{eqnarray}
\label{eq:Y}
Y(\omega)&=&\frac 1R\left[1- {4\pi K}{\cal G}(\omega)\right], 
\\
{\cal G}(\omega)&=&\frac{\lambda^2}{-i\omega}\left[G_{\sin 2\pi n,\sin2\pi n}(\omega)-G_{\sin 2\pi n,\sin2\pi n}(\omega=0)\right].
\nonumber
\end{eqnarray}
Let us emphasize that Eq.~\eqref{eq:Y} is valid at any $\lambda$.

Equation \eqref{eq:Y}  allow establishing a relation between the admittance of the boundary sine-Gordon model and the conductance across an impurity in a Luttinger liquid~\cite{Kane1992,Matveev1993}. Indeed, the first term in Hamiltonian~\eqref{eq:BSG2} may alternatively describe one-dimensional interacting fermionic leads. In that context, the parameter $K$ characterize whether the interactions between fermions are repulsive (at $K<\frac12$) or attractive (at $K >\frac12$). Using linear response, we then find that  $e^2{\cal G}(\omega)$ in Eq.~\eqref{eq:Y} can be interpreted as the dynamical conductance of an electronic junction with tunnel current operator ${\cal I}=e\lambda\sin 2\pi (n+{\cal N})$ under voltage bias ${\cal V}=2\pi \dot{\cal N}/e$.

\subsection{Perturbation theory in $\lambda$ at $K>\frac 12$}

It is straightforward to evaluate the admittance up to order $\lambda^2$ with Eq.~\eqref{eq:Y}. For this we just need to insert in it the low-frequency result for the Green function $G_{\sin2\pi n,\sin 2\pi n}(\omega)$ evaluated at $\lambda=0$,
\begin{equation}
\label{eq:G-sinsin}
G_{\sin 2\pi n,\sin2\pi n}(\omega)=\frac{-i\pi e^{-i2\pi K}}{2\Gamma(4K)\cos(2\pi K)}\frac {\omega^{4K-1}}{\omega_0^{4K}},\quad \omega\ll \omega_0,
\end{equation}
In particular, at $K>\frac 12$, $G_{\sin 2\pi n,\sin2\pi n}(\omega=0)=0$. The correction to $Y'(\omega)=1/R$ remains small at any frequency~\cite{Aslangul1987,Kane1992}. 
It was noticed in a related context~\cite{Aslangul1987} that $Y''(\omega)$ is positive (inductive-like) at $\frac 12<K<\frac 34$ and diverges at $K=\frac 34$. The divergence signals an insufficiency of the low-energy model \eqref{eq:bsG}. Indeed, at $K>\frac34$ the power-law in the $\omega$-dependence associated with the quantum response exceeds 1. This signals that the quantum response is superseded by the classical one [not accounted for in Eq.~\eqref{eq:bsG}], $Y''(\omega)\simeq \omega L/R$, which also remains small up to a narrow vicinity of $\omega_0$. Overall, the correction to $\delta(\omega)=\pi/2$ remains small at any frequency up to $\omega_0$, as announced in the main text.

Let us remind here that perturbation theory is also fruitful to discuss the high-frequency regime at $K<\frac12$, see the main text.

\subsection{Dual model near the IR fixed point}

At $K<\frac 12$, the boundary condition of the free bosonic field at the position of the impurity (the junction) changes from $\partial_x\theta(0)=0$ to $\partial_x\varphi(0)=0$ (i.e., from von Neumann to Dirichlet condition as far as the $\theta$-field is concerned) between the UV fixed point corresponding to $\lambda=0$ in Eq.~(8) of the main text, and the IR fixed point corresponding to $\tilde \lambda=0$ in Eq.~(14) of the main text. To address the vicinity of the IR fixed point, it is convenient to introduce another set of bosonic annihilation/creation operators, $b_q=(x_q+ip_q)/\sqrt{2}$ and $b_q^\dagger=(x_q-ip_q)/\sqrt{2}$, such that Eq.~(14) of the main text reads:
\begin{eqnarray}
\label{eq:BSG-dual}
H_{\rm dual}&=&\sum_q\frac{\omega_q}2\left(x_q^2+p_q^2\right)-\tilde \lambda\cos\varphi -\varphi \dot{\cal N},\quad \\
\nonumber 
\varphi&=&\sum_q\sqrt{\frac{2\Delta}{K\omega_q}}p_q.
\end{eqnarray}
By linear response, the admittance is 
\begin{equation}
Y(\omega)=\frac 1R\left[1+\frac K\pi G_{\dot \varphi,\varphi}(\omega)\right],
\end{equation}
where the Green function $G_{\dot \varphi,\varphi}(\omega)$ is evaluated with Eq.~\eqref{eq:BSG-dual} at ${\cal N}=0$.
Similar to Sec.~\ref{sec:EOM}, by using the equations of motion for $x_q,p_q$ we find various identities between Green functions and equivalent expressions for the admittance,
\begin{eqnarray}
\label{eq:Ybis}
Y(\omega)&=&-\frac{\tilde \lambda}RG_{\varphi,\sin\varphi}(\omega)=4e^2\tilde {\cal G}(\omega), 
\\
\tilde {\cal G}(\omega)&=&\frac{\tilde \lambda^2}{-i\omega}\left[G_{\sin \varphi,\sin\varphi}(\omega)-G_{\sin \varphi,\sin\varphi}(\omega=0)\right].
\nonumber
\end{eqnarray}
Here we used $\langle \cos \varphi\rangle =-\tilde\lambda G_{\sin \varphi,\sin\varphi}(\omega=0)$.
Let us emphasize that all relations above are valid at any $\tilde \lambda$.

The term $\propto \tilde \lambda$ is an irrelevant perturbation near the IR fixed point. Therefore we may use the expression for $G_{\sin \varphi,\sin\varphi}(\omega)$ evaluated at $\tilde \lambda=0$ to evaluate the admittance up to order $\tilde \lambda^2$ down to the lowest frequencies. Noting that $G_{\sin \varphi,\sin\varphi}(\omega)$ can be read off the r.h.s.~of Eq.~\eqref{eq:G-sinsin} after the substitution $K\to 1/4K$, and using Eq.~(3) in the main text to convert the admittance into scattering phase shift, we readily find the low-frequency asymptote, Eq.~(16) in the main text.

Actually, as discussed in the main text, a series of additional perturbations, 
\begin{equation}
V=\sum_{n>0}c_{2n}(\partial_x\theta(0))^{2n},\quad \partial_x\theta(0)=\sum_q\frac{\sqrt{K\Delta\omega_q}}v(b_q+b^\dagger _q).
\end{equation}
need to be added to the dual Hamiltonian \eqref{eq:BSG-dual}, 
\begin{equation}
H_{sG}=H_{\rm dual}+V,
\end{equation}
in order to address the low-frequency response functions at $K<\frac13$. The leading term $\propto c_2$ is quadratic in the $\theta$-field. Therefore it only contributes to elastic scattering. It yields a perturbative-in-$c_2$ contribution to the boson $T$-matrix, $T_{qq'}=c_2K\Delta\sqrt{\omega_q\omega_q'}/v^2$ and, consequently, a linear-in-$\omega$ contribution to the elastic phase shift, $\delta(\omega)=-\pi c_2K\omega/v^2\ll 1$ with $c_2<0$. This contribution with $\pi c_2K/v^2=-1/\beta(K)\omega_\star$ corresponds to the first term in the r.h.s.~of Eq.~(17)  in the main text. To find the dependence of $c_2$ on the circuit parameters, we note that such a linear-in-$\omega$ phase shift  corresponds to an additive contribution to the boson density of states, \begin{equation}
\delta \nu=\frac 1\pi\frac{d\delta(\omega)}{d\omega}=c_2K/v^2. 
\end{equation}Therefore the boundary induces an additive correction to the specific heat, which takes a simple form in one dimension, $\delta C/T=(\pi^2/3)\delta \nu=\pi/3\beta(K)\omega_\star$.  We benefited from an exact result for $\delta C/T$ given in Eq.~(5.23) of Ref.~\cite{Bazhanov1997} to find the expression for $\beta(K)$ at $K<\frac 13$, which appears in Eq.~(17) in the main text. Alternatively we might have used the expression for $c_2$ given in Eq.~(3.21) of Ref.~\cite{Lesage1999}. (From the comparison with \cite{Bazhanov1997}, we suspect that the exponent of the factor $2\sin\pi g$ is actually twice larger than indicated there.)

The quartic term $\propto c_4$ contributes to the inelastic scattering cross-section,
\begin{eqnarray}
\sigma_{\rm in}(\omega)&=&\frac {4\pi^2}{3!}\frac{c_4^2K^4}{v^8}\int_0^\omega \!\!\!d\omega_1\int _0^{\omega-\omega_1}\!\!\!\!\!\!\!\!\!d\omega_2 \omega\omega_1\omega_2(\omega-\omega_1-\omega_2)
\nonumber\\
&=&\frac {\pi^2}{180}\frac{c_4^2K^4\omega^6}{v^8}.
\end{eqnarray}
This inelastic contribution becomes dominant at $K<\frac 18$, as stated in footnote [32] in the main text. Taking into account an aforementioned modification of Eq.~(3.21) in Ref.~\cite{Lesage1999}, we find
\begin{equation}
\delta''(\omega)=\frac{\sigma_{\rm in}(\omega)}{4}=\eta(K)\left(\frac\omega{\omega_\star}\right)^6,
\end{equation}
where, in particular, $\eta(K)\propto K^2$ at $K\ll 1$.

\subsection{Toulouse point, $K=\frac 14$}

At $K=\frac14$ (Toulouse point), we use the mapping to a fermionic field, 
\begin{equation}
\label{eq:fermion}
\psi(x)=\sqrt{\frac{\omega_0}{2\pi v}} \gamma e^{2i\pi n(x)}=\frac1{\sqrt{2L}}\sum_k\psi_ke^{ikx},
\end{equation}
where $\gamma$ is a Majorana fermion ($\gamma^2=1$), to express the Hamiltonian \eqref{eq:BSG2} at ${\cal N}=0$ as~\cite{Kane1992,Guinea1985b}
\begin{equation}
H_{\rm sG}=\sum_k  vk \psi^\dagger_k\psi_k+\frac{\lambda}2\sqrt{\frac{2\pi v}{\omega_0}} \left[\psi(0)-\psi^\dagger(0)\right]\gamma.
\end{equation}
As the Hamiltonian is quadratic, the dynamical conductance $e^2{\cal G}(\omega)$ associated with it is easily obtained. Inserting the result in Eq.~\eqref{eq:Y}, one obtains
\begin{equation}
\label{eq:Toulouse}
RY(\omega)=1+\frac{\omega_c}{i\omega}\ln\left(1-\frac{i\omega}{\omega_c}\right), \quad \omega_c=\frac{\pi \lambda^2}{\omega_0}.
\end{equation}
Equation (20) in the main text is readily obtained from this result using the relation between the admittance and scattering phase given in Eq.~(3) in the main text.

\subsection{Weakly interacting fermions, $\frac12-K\ll 1$}
Another mapping is possible at the free-fermion point, $K=\frac12$~\cite{Guinea1985}, by introducing two fermionic fields, $\psi_R$ and $\psi_L$, as in Eq.~\eqref{eq:fermion}. Then, Eq.~\eqref{eq:BSG2} at ${\cal N}=0$ transforms into a conventional tunnel Hamiltonian,
\begin{equation}
\label{eq:free}
H_{sG}=\sum_{k,\sigma=L,R}vk\psi^\dagger_{k\sigma}\psi_{k\sigma}+\frac{\pi v \lambda}{\omega_0} \left[\psi^\dagger_L(0)\psi_R(0)+\mathrm{H.c.}\right].
\end{equation}
The $\omega$-dependence of the tunnel conductance vanishes,  
\begin{equation}
{\cal G}=\frac1{2\pi}{\cal T},\qquad {\cal T}=\frac {(\pi\lambda/\omega_0)^2}{1+(\pi\lambda/\omega_0)^2}.
\end{equation}
Thus $RY(\omega)=1-{\cal T}\equiv {\cal R}$ along the critical line $K=\frac12$. 

The mapping of Eq.~\eqref{eq:free} can be extended to arbitrary $K$ by considering that the fermions in the leads are interacting. In particular, we may use the renormalization of the transmission amplitude found at weakly repulsive interaction, $0<\frac12-K\ll 1$, in Ref.~\cite{Matveev1993} to find 
\begin{equation}
\label{eq:MYG}
RY(\omega)=\frac{1}{1+(1-i2\pi \delta K)(\omega/\omega_\star)^{4\delta K}},\quad \delta K=K-\frac12.
\end{equation}
Here $\omega_\star=\omega_0(\pi\lambda/\omega_0)^{1/(1-2K)}$, in agreement with Eq.~(13) of the main text at $K\to \frac 12$. Note that the vanishing imaginary part in Eq.~\eqref{eq:MYG} is found from the matching of $\delta(\omega)$ given in Eq.~(21) in the main text [and obtained from Eq.~\eqref{eq:MYG} and Eq.~(3) in the main text], with its high- and low-frequency asymptotes, Eqs.~(12) and (16) in the main text,  respectively. 

From Eq.~\eqref{eq:MYG} we find that $Y(\omega)\to 1/2R$ at $\omega=\omega_\star$ and $K\to \frac12-0$. Thus $\sigma_{\rm in}(\omega)\to 1$: scattering is fully inelastic. This result matches Eq.~\eqref{eq:free} found at $K=\frac 12$ when one considers the scaling limit, $\omega_0\to \infty$ with fixed $\omega_\star$, such that $\lambda/\omega_0\simeq (\omega_\star/\omega_0)^{-2\delta K}\to 1$ and ${\cal T},{\cal R}\to \frac 12$.